\RequirePackage{fix-cm}
\documentclass{svjour3}                     
\smartqed  
\usepackage{graphicx}
\usepackage{color}
\usepackage{mathptmx}      
\begin{document}

\title{
Note on the perihelion/periastron advance
due to cosmological constant
}


\author{Hideyoshi Arakida
}


\institute{H. Arakida \at
Graduate School of Education, Iwate University,\\
3-18-33, Ueda, Morioka, Iwate 020-8550, Japan
              \email{arakida@iwate-u.ac.jp}           
}


\maketitle

\begin{abstract}
 We will comment on the perihelion/periastron advance of celestial 
 bodies due to the cosmological constant $\Lambda$. It is well
 known that the cosmological constant $\Lambda$ causes the 
 perihelion/periastron shift; however, there seems to still exist
 a discrepancy among the various derived precession 
 formulae. We will point out that the expression 
 $\Delta \omega_{\Lambda} = (\pi c^2 \Lambda a^3/(GM))\sqrt{1 - e^2}$
 is the general formula for any orbital eccentricity $e$
 and the expression
 $\Delta \omega_{\Lambda} = (\pi c^2 \Lambda a^3/(GM))(1 - e^2)^3$
 comes from the nearly circular ($e \ll 1$) approximation.
\end{abstract}
\section{Introduction}
The 2011 Nobel Prize in physics was awarded for the discovery of 
the accelerating expansion of the Universe \cite{riess1998,perlmutter1999}. 
According to the present literature, theory that is most suitable 
for explaining this phenomenon is that of the cosmological constant 
$\Lambda$, generally referred to as dark energy. Although the
cosmological constant or dark energy can be considered to account 
for the current expansion of the Universe in the simple terms, 
the details are still far from clear. Meanwhile, several 
attempts have also been made for explaining this accelerating expansion 
without dark energy
\cite{bekenstein2004,capozziello2008,nicolis2009,harova2009a,harova2009b,harova2009c,sotiriou2010,defelice2010}.

To verify the existence of the cosmological constant/dark energy and/or
the potentiality of alternative gravitational theories, 
it is most natural to investigate 
their properties in the context of classical tests of general
relativity, i.e.,
the perihelion/periastron advance of celestial bodies and the 
deflection of light. To this end, we must elucidate the difference 
in such effects among the cosmological constant and the alternative 
gravitational theories. Therefore, it is important to derive a
rigorous formula for such effects resulting from the cosmological
constant $\Lambda$. Recently, the effect of the cosmological constant 
on the bending of light rays was intensively investigated and 
discussed by many authors; the detailed reports can be found in 
\cite{rindler2007,ishak2010,lake2002,park2008,kp2008,sph2010,bhadra2010,miraghaei2010,biressa2011,arakida2012}
and the references therein.

The influence of the cosmological constant on the orbital 
motion is a historical problem (see section 45 of \cite{eddington}). 
In principle, it is clear that the cosmological constant contributes to
the perihelion/periastrion advance, although this contribution is too
small to detect because of 
$\Lambda \approx 10^{-52} ~{\rm m^{-2}}$. 
However, there is 
a discrepancy in the obtained precession formulae, in terms of 
the difference in the eccentricity dependence $(1 - e^2)^3$ 
\cite{eddington,rindler2006,miraghaei2010} and
$\sqrt{1 - e^2}$ \cite{kerr2003,iorio2008}. 
The contribution of cosmological constant on the orbital 
dynamics is also investigated in other papers 
\cite{islam1983,ct1998,kw2003,iorio2006,js2006,kkl2006,sj2006,js2007,dumin2008,martin2012},
and constraints on the cosmological constant from the Sun's motion 
through the Milky Way is recently discussed \cite{iorio2010}.

In fact, this issue has been resolved in \cite{adkins2007} 
using the general framework of radial/central force perturbation.
Subsequently, it was shown that the physical interpretation of 
the formula derived in \cite{adkins2007} is the 
precession of Hamilton's vector \cite{chashchina2008}. A similar 
problem as that in \cite{adkins2007} was treated in terms of 
the Gaussian planetary equation \cite{ruggiero2010}, 
and Multiple Scales Method \cite{mpejic}.
In \cite{adkins2007,ruggiero2010,mpejic}, it was shown that the correct 
eccentricity dependence is $\sqrt{1 - e^2}$ as previously shown
in \cite{kerr2003,iorio2008}. However, in spite of 
these facts, some confusion still seems to exist 
(Refer to e.g., \cite{miraghaei2010}). Therefore, the purpose of
this short note is to arrange the discussions on the perihelion/periastron
advance due to the cosmological constant $\Lambda$.
\section{Perihelion/periastron advance
due to cosmological constant}
\subsection{Analytical approach by direct integration of
perturbation potential/force}
In order to examine the additional perihelion/periastron advance
due to $\Lambda$, let us assume the spacetime to be spherically
symmetric; without loss of generality, we write the metric as
\begin{eqnarray}
 ds^2 = - \left[1 - \frac{r_g}{r} + W(r)\right]c^2 dt^2
  + \left[
     1 - \frac{r_g}{r} + W(r)
    \right]^{-1}dr^2 + r^2 d\Omega^2,
\end{eqnarray}
where $r_g \equiv 2GM/c^2$, 
$d\Omega^2 = d\theta^2 + \sin^2 \theta d\phi^2$ and 
$W(r)$ is an additional term that is a function of $r$ and
generally depends on the gravitational theories. Then, the 
geodesic equation of the test particle with mass $m$ becomes
\begin{eqnarray}
 & &
 \left(\frac{dr}{d\tau}\right)^2 + U(r) = 
  \left(\frac{E}{mc}\right)^2,\\
 & &
  U(r) = c^2 \left(1 + \frac{L^2}{c^2 m^2 r^2}\right)
  \left[
   1 - \frac{r_g}{r} + W(r)
  \right].
\end{eqnarray}
Here, $\tau$ is the proper time, and
$E$ and $L$ are two constants of motion: 
the total energy and 
angular momentum, respectively. According to the standard approach,
we define the Newtonian potential $\Phi (r)$ as
\begin{eqnarray}
 \Phi (r) \equiv
  \frac{1}{2}\lim_{c^2 \rightarrow \infty}
  \left[U(r) - c^2\right]
  = -\frac{GM}{r} + \frac{{\cal L}^2}{2 r^2} + \frac{1}{2}W(r)c^2,
\end{eqnarray}
where we set ${\cal L} = L/m$ and let ${\cal E} = (1/2)(E/mc)^2$; 
then, we have,
\begin{eqnarray}
 \frac{1}{2} \left(\frac{dr}{d\tau}\right)^2 + \Phi(r) = {\cal E}.
\label{eq:eom1}
\end{eqnarray}
If we rewrite $V(r) \equiv (1/2)W(r) c^2$ 
and $\tau \approx t$ ($t$ is coordinate time), 
Eq. (\ref{eq:eom1}) becomes completely equivalent to Eq. (15) of 
\cite{adkins2007}. Thus, henceforth, we follow the approach 
employed in \cite{adkins2007}.

The perihelion/periastron shift $\Delta \theta_p$ 
due to the perturbation potential $V(r)$ 
with respect to Newtonian potential $-GM/r$
is expressed as (see Eq. (30) in \cite{adkins2007})
\begin{eqnarray}
 \Delta \theta_p
 &\equiv& \frac{-2 \ell}{GM e^2}\int^{1}_{-1}
 \frac{zdz}{\sqrt{1 - z^2}}
 \frac{dV(z)}{dz},
\label{eq:peri-formula1} 
\end{eqnarray}
in which $r = \ell/(1 + ez)$ and $\ell = {\cal L}^2/(GM) = a(1 - e^2)$,
$a$ being the semi-major axis of the orbit.
It should be noted that in \cite{adkins2007},
$V(r)$, or equivalently $V(z)$, is defined including the mass of
the orbiting particle $m$  (refer to  Eq. (15) in \cite{adkins2007}).
However, in our case, $V(r)$ or $V(z)$ does not contain $m$;
hence, hereafter, $m$ does not appear in the precession formula.

We are interested in the power-law perturbing potentials of the type
$V(r) = \alpha_n r^n$. Noting the relation,
\begin{eqnarray}
V(r) = - \alpha_{- (n + 1)} r^{- (n + 1)}
= - \alpha_{- (n + 1)} (1 + ez)^{n + 1}/\ell^{n + 1},
\end{eqnarray}
we have the precession formula,
\begin{eqnarray}
 \Delta_p (- (n + 1)) =
  - \frac{2 \alpha_{- (n + 1)} (n + 1)}{GM \ell^n e}
  \int^{1}_{-1}\frac{z (1 + ez)^n}{\sqrt{1 - z^2}}dz.
\label{eq:peri-formula2}
\end{eqnarray}
For discussing the effect of cosmological constant $\Lambda$, the
Schwarzschild--de Sitter or Kottler metric \cite{kottler1918},
\begin{eqnarray}
  ds^2 = 
  - \left(1 - \frac{r_g}{r} - \frac{\Lambda}{3}r^2\right)
  c^2dt^2 + 
  \left(1 - \frac{r_g}{r} - \frac{\Lambda}{3}r^2\right)^{-1}dr^2
  + r^2 d\Omega^2
  \label{eq:sdsk}
\end{eqnarray}
is widely used and $r_g/r$ term causes famous general relativistic
perihelion/periastron shift. In the case of perihelion/periastron 
advance due to $\Lambda$, we consider $W (r) = - (1/3)\Lambda r^2$ 
and then $V(r) = - (1/6)\Lambda c^2 r^2$. 
It should be noted in this case, $- (n + 1) = 2$; thus, $n = -3$ and
$\alpha_{2} = - (1/6)\Lambda c^2$.

The integral part in Eq. (\ref{eq:peri-formula2}) is
\begin{eqnarray}
\int^1_{-1} \frac{z (1 + e z)^{-3}}{\sqrt{1 - z^2}}dz
=
-\frac{3\pi}{2} \frac{e \sqrt{1-e^2}}{(1-e^2)^3}.
\label{eq:peri-formula3}
\end{eqnarray}
Therefore, from Eqs. (\ref{eq:peri-formula2}) and 
(\ref{eq:peri-formula3}), the perihelion/periastron advance 
due to $\Lambda$ is given by the following equation
(again, note $\ell = a (1 - e^2)$
and $n = -3$)
\begin{eqnarray}
\Delta \omega_{\Lambda} = \frac{\pi c^2 \Lambda a^3}{GM}
 \sqrt{1 - e^2}.
\label{eq:adv1}
\end{eqnarray}
Eq. (\ref{eq:adv1}) is in agreement with the results of
\cite{kerr2003,iorio2008,ruggiero2010}.

Let us summarize the outline of above derivation. We start
from the SdS/Kottler metric Eq. (\ref{eq:sdsk}) and its 
``Newtonian approximation (weak field approximation)''. 
Then the perturbation potential $U_{\Lambda} = -(1/6)\Lambda r^2$ 
or perturbation force $F_{\Lambda} = (1/3)\Lambda r$ is determined. 
The result, Eq. (\ref{eq:adv1}), is obtained exactly by 
``the direct integration of given perturbation potential or force'' 
e.g. Eq. (\ref{eq:peri-formula3}), with the help of the knowledge on
Gaussian hypergeometric function $_2 F_1 (\alpha, \beta, \gamma;~x)$,
see Eq. (37) in \cite{adkins2007}.
We mention that Eq. (\ref{eq:peri-formula3}) can be shortly
calculated by the variable transformation,
\begin{eqnarray}
 z = \frac{\cos \xi - e}{1 - e\cos \xi},
\end{eqnarray}
which produces the integral
\begin{eqnarray}
 \frac{\sqrt{1 - e^2}}{(1 - e^2)^3}
  \int^{\pi}_0 (\cos \xi - e)(1 - e\cos \xi)d\xi
  =
  -\frac{3\pi}{2} \frac{e \sqrt{1-e^2}}{(1-e^2)^3}.
\end{eqnarray}
It is worthy to note when integrating Eq. (\ref{eq:peri-formula3}),
the integral part is not expanded with respect to the orbital
eccentricity $e$.
\subsection{Standard perturbation method}
The perihelion/periastron advance due to $\Lambda$ is considered 
by means of the standard perturbation method. Here let us discuss 
this approach according to \cite{rindler2006}.

Adopting SdS/Kottler metric Eq. (\ref{eq:sdsk}), we begin with 
the second-order geodesic equation for 
time-like world-line,
\begin{eqnarray}
 \frac{d^2 u}{d\phi^2} + u - \frac{GM}{L^2_z}=
  \frac{3GM}{c^2}u^2
  - \frac{\Lambda c^2}{3L^2_z u^3},\quad
  u = \frac{1}{r},
  \label{eq:geo2}
\end{eqnarray}
here $L_z$ is the $z$ component of orbital angular momentum
\footnote{In \cite{rindler2006}, the $z$ component of angular
 momentum is expressed by $h$ instead of $L_z$.},
and right-hand side of (\ref{eq:geo2}) can be considered as the
perturbation to the Keplerian motion. The Keplerian motion
is described by
\begin{eqnarray}
 \frac{d^2 u}{d\phi^2} + u - \frac{GM}{L^2_z} = 0
\end{eqnarray}
and its solution is
\begin{eqnarray}
 \frac{1}{r} = u = \frac{GM}{L^2_z}(1 + e \cos \phi).
  \label{eq:adv3}
\end{eqnarray}
Since it is well-known that the first term in right-hand side of
Eq. (\ref{eq:geo2}) causes famous general relativistic precession
formula, $\Delta \omega_{\rm gr} = 6\pi GM/(ac^2 (1 - e^2))$, and
Eq. (\ref{eq:geo2}) is linear differential equation, 
then we concentrate on the perturbation due to the cosmological
constant $\Lambda$;
\begin{eqnarray}
 \frac{d^2 u}{d\phi^2} + u - \frac{GM}{L^2_z}=
  - \frac{\Lambda c^2}{3L^2_z u^3}.
  \label{eq:geo3}
\end{eqnarray}
Inserting Eq. (\ref{eq:adv3}) into right-hand side of 
Eq. (\ref{eq:geo3}), we have
\begin{eqnarray}
 - \frac{\Lambda c^2}{3L^2_z u^3}
  = 
  - \frac{\Lambda c^2 L^4_z}{3(GM)^3}
  \frac{1}{(1 + e \cos\phi)^3}.
\end{eqnarray}
Here, it is worthy to mention when obtaining Eq. (\ref{eq:adv1}),
the part $1/(1 + e \cos \phi)^3$ is exactly integrated with changing 
the variable $\cos\phi$ with $z$ (see Eq. (\ref{eq:peri-formula3})).
However, in the standard literature such as \cite{rindler2006}, 
this part is expanded with respect to the orbital eccentricity $e$,
\begin{eqnarray}
 \frac{1}{(1 + e \cos\phi)^3} = 1 - 3 e\cos\phi + 6e^2 \cos^2\phi
  + {\cal O}(e^3),
\end{eqnarray}
and only $e\cos\phi$ term is retained in \cite{rindler2006}.
Comparing with the derivation process of general relativistic
precession formula, the precession formula due to $\Lambda$ is 
derived as (see Eq. (14.25) in \cite{rindler2006})
\begin{eqnarray}
 \Delta \omega_{\Lambda} \approx \frac{\pi c^2 \Lambda L^6_z}{(GM)^4}.
  \label{eq:rindler1}
\end{eqnarray}
It should be emphasized here although the expansion itself is 
``truncated at the first order in $e$'', the orbital angular
momentum $L_z$ is given by
\begin{eqnarray}
 L_z = \sqrt{GM\ell} = \sqrt{GMa (1 - e^2)}
  \label{eq:rindler2}
\end{eqnarray}
{\it then the quadratic terms in orbital eccentricity $e$ is 
remained in the final result due to $L_z$ but these terms are 
not cared in the lowest order perturbation method considered}.

Inserting Eq. (\ref{eq:rindler2}) into Eq. (\ref{eq:rindler1}), 
it is found
\begin{equation}
\Delta \omega_{\Lambda} = 
 \frac{\pi c^2 \Lambda a^3}{GM} (1 - e^2)^3,
\label{eq:adv2}
\end{equation}
which is also same results of \cite{eddington,miraghaei2010}.

If we expand the integral part in Eq. (\ref{eq:peri-formula3}) 
up to the second order in $e$ and integrate, then we have
\begin{eqnarray}
\int^1_{-1} \frac{z (1 + e z)^{-3}}
 {\sqrt{1 - z^2}}dz
\simeq
\int^1_{-1} \frac{z (1 - 3ez + 6 e^2 z^2)}
{\sqrt{1 - z^2}}dz
=
- \frac{3\pi}{2} e,
\label{eq:peri-formula4}
\end{eqnarray}
and combining with Eq. (\ref{eq:peri-formula2}), 
Eq. (\ref{eq:adv2}) is recovered.

\begin{figure}[h]
\begin{center}
 \includegraphics[scale=0.8]{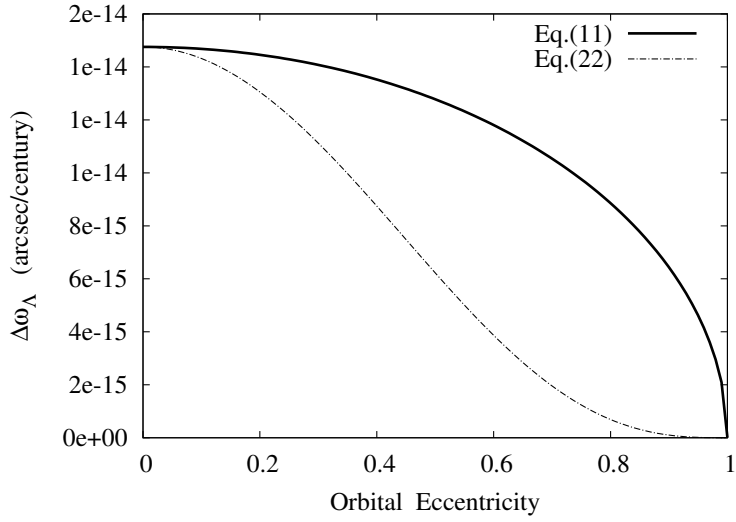}
\end{center}
 \caption{The eccentricity dependence of Eqs. (\ref{eq:adv1}) and 
(\ref{eq:adv2}). In this plot, we set 
$\Lambda \approx 10^{-52}~{\rm m^{-2}}$,
$M = 2.0 \times 10^{30}~{\rm kg}$, 
$a = 1.5 \times 10^{11}~{\rm m}$.
\label{fig1}}
\end{figure}

The eccentricity dependence of Eqs. (\ref{eq:adv1}) and 
(\ref{eq:adv2}) is plotted in Fig. \ref{fig1}. It is clear
that Eqs. (\ref{eq:adv1}) and (\ref{eq:adv2}) behave quite
differently as the orbital eccentricity becomes large.
\section{Summary}
In this short note, we commented on the perihelion/periastron advance 
of celestial bodies due to the cosmological constant $\Lambda$.
As we stated before, our prescription is based on \cite{adkins2007},
in which the problem of perihelion/periastron shift due to the 
cosmological constant $\Lambda$ has practically been resolved. 
However, we hope that this note contributes to clear up the
confusion about the perihelion/periastron advance due to the
cosmological constant $\Lambda$, and is helpful in discussing the 
probability of alternative theories of gravitation 
since $\Lambda$-like terms arise also in various theoretical 
contexts such as $f(T)$ \cite{is2012} and $f(R)$ gravity \cite{ri2007}.

Just before closing this letter, 
we would mention the following issue; 
for high orbital eccentricity $e$ and large central mass $M$,
the general relativistic value of perihelion/periastron 
advance becomes large, while {\it Eqs. (\ref{eq:adv1}) and
(\ref{eq:adv2}) tend to zero} (see again Fig. \ref{fig1}).
This property in Eqs. (\ref{eq:adv1}) and (\ref{eq:adv2}) seems to
be physically counter-intuitive. As a possibility, this
counter-intuitive feature may have roots in the fact that essentially,
Schwarzschild--de Sitter/Kottler metric does not become 
asymptotically flat spacetime. However the derivation of Eqs. 
(\ref{eq:adv1}) and (\ref{eq:adv2}) may assume implicitly that 
the background metric is Minkowskian. Therefore, the problem on 
perihelion/periastron advance due to cosmological constant $\Lambda$
should be investigated and discussed further in
a careful manner in the context of real expanding universe 
(see, for example, \cite{bbp2001}, and similar topic was recently
discussed by \cite{iorio2012}).

\begin{acknowledgements}
We would like to acknowledge two anonymous referees for 
reading our manuscript carefully and for giving fruitful
comments and suggestions, which significantly improved the
quality of the manuscript. We also thank M. Kasai for the
fruitful discussions.
\end{acknowledgements}

\begin{thebibliography}{}
 \bibitem{riess1998} A. G. Riess, {\it et al.}, Astron. J., 
	 116, 1009 (1998).

\bibitem{perlmutter1999}S. Perlmutter, {\it et al.}, 
	Astrophys. J., 517, 565 (1999).

\bibitem{bekenstein2004}J. D. Bekenstein, 
	Phys. Rev. D, 70, id. 083509 (2004).

\bibitem{capozziello2008}S. Capozziello, M. Francaviglia,
	Gen. Relativ. Gravit., 40, 357 (2008).

\bibitem{nicolis2009}A. Nicolis, R. Rattazzi, E. Trincherini,
	Phys. Rev. D, 79, id. 064036 (2009).

\bibitem{harova2009a}P. Ho\v{r}ava, 
	J. High Energy Phys., 03, 020 (2009).

\bibitem{harova2009b}P. Ho\v{r}ava, 
	Phys. Rev. D, 79, id. 084008 (2009).

\bibitem{harova2009c}P. Ho\v{r}ava, 
	Phys. Rev. Lett., 102, id. 161301 (2009).

\bibitem{sotiriou2010}T. Sotiriou, V. Faraoni,
	Rev. Mod. Phys., 82, 451 (2010).

\bibitem{defelice2010}A. De Felice, S. Tsujikawa, 
	Living Reviews in Relativity, 13, no. 3 (2010).

\bibitem{rindler2007} W. Rindler, M. Ishak, 
	Phys. Rev. D, 76, id. 043006 (2007).

\bibitem{ishak2010} M. Ishak, W. Rindler, 
	Gen. Relativ. Gravit., 42,  2247 (2010).

\bibitem{lake2002} K. Lake, 
	Phys. Rev. D, 65, id. 087301 (2002). 

\bibitem{park2008} M. Park, 
	Phys. Rev. D, 78, id. 023014 (2008).

\bibitem{kp2008} I. B. Khriplovich, A. A. Pomeransky,
	Int. J. Mod. Phys. D, 17, 2255 (2008).

\bibitem{sph2010} F. Simpson, J. A. Peacock, A. F. Heavens,
	Mon. Not. R. Astron. Soc., 402, 2009 (2010).

\bibitem{bhadra2010}A. Bhadra, S.  Biswas, K. Sarkar, 
	Phys. Rev. D, 82, id 063003 (2010).

\bibitem{miraghaei2010} H. Miraghaei, M. Nouri-Zonoz, 
	Gen. Relativ. Gravit., 42, 2947 (2010).

\bibitem{biressa2011} T. Biressa, J. A. de Freitas Pacheco, 
	Gen. Relativ. Gravit., 43, 2649 (2011).

\bibitem{arakida2012} H. Arakida, M. Kasai,
	Phys. Rev. D, 85, id 023006 (2012).

\bibitem{eddington} A. S. Eddington, 
	The Mathematical Theory of Relativity,
	Cambridge University Press, Cambridge (1923).

\bibitem{rindler2006} W. Rindler, 
	Relativity: Special, General, and
	Cosmological (2nd Ed.), Oxford Univ. Press, 
	New York (2006).

\bibitem{kerr2003} A. W. Kerr, J. C. Hauck, B. Mashhoon,
	Class. Quant. Grav., 20, 2727 (2003).

\bibitem{iorio2008} L. Iorio, 
	Adv. Astron., id. 268647 (2008).

 \bibitem{islam1983}J. N. Islam, 
	 Phys. Lett. A, 97, 239 (1983).

\bibitem{ct1998}J. F. Cardona, and J. M Tejeiro, 
	Astrophys. J., 493, 52 (1998).

\bibitem{kw2003} G. V. Kraniotis and S. B. Whitehouse,
	Class. Quant. Grav., 20, 4817 (2003).

\bibitem{iorio2006} L. Iorio, Int. J. Mod. Phys. D, 15, 
	473 (2006).

\bibitem{js2006}Ph. Jetzer and M. Sereno, 
	Phys. Rev. D, 73, id 044015 (6 pages) (2006).

\bibitem{kkl2006}V. Kagramanova, J. Kunz, and C. L\"ammerzahl, 
	Phys. Lett. B, 634, 465 (2006).

\bibitem{sj2006}M. Sereno and Ph. Jetzer, 
	Phys. Rev. D, 73, id 063004 (5 pages) (2006).

 \bibitem{js2007}Ph. Jetzer, and M. Sereno, 
	 Il Nuovo Cimento B, 122, 489 (2007).

\bibitem{dumin2008} Y. V. Dumin,, Proc. MG11 Meeting on General Relativity,
(Edited by Hagen Kleinert, Robert T Jantzen, Remo Ruffini).
	World Scientific Publishing, 1752 (2008).

\bibitem{martin2012}J. Martin, Comptes Rendus de l'Academie des
	Sciences, 13, 566 (2012) [arXiv:1205.3365].

\bibitem{iorio2010} L. Iorio, MNRAS, 403, 1469 (2010).

\bibitem{adkins2007}G. S. Adkins, J. McDonnell,
	Phys. Rev. D, 75, id. 082001 (2007).

\bibitem{chashchina2008} O. I. Chashchina, Z. K. Silagadze, 
	Phys. Rev. D, 77, id. 107502 (2008).

\bibitem{ruggiero2010} M. L. Ruggiero, arXiv:1010.2114 (2010).

\bibitem{mpejic} M. Pejic, Calculating Perihelion Precession
Using The Multiple Scales Method,
http://math.berkeley.edu/$^{\sim}$mpejic/pdfdocuments/PerihelionPrecession.pdf

\bibitem{kottler1918} F. Kottler, Annalen Phys., 361, 401 (1918).

\bibitem{is2012}L. Iorio and E. N. Saridakis, 
	MNRAS, 427, 1555 (2012)  [arXiv1203.5781I].

\bibitem{ri2007}M.L. Ruggiero and L. Iorio, 
	JCAP, 01, id 010 (2007).

\bibitem{bbp2001} B. Bolen, L. Bombelli and R. Puzio, 
	Class. Quant. Grav., 18, 1173 (2001) [gr-qc/0009018].

\bibitem{iorio2012} L. Iorio, MNRAS, in press (2012) 
	[arXiv:1208.1523].

%
\end{thebibliography}


\end{document}